\documentclass[onecolumn,showpacs,amsmath,amssymb,prd,nofootinbib]{revtex4-2}
\usepackage{graphicx}
\usepackage{epsfig}
\usepackage{enumerate}
\def\beq{\begin{equation}}
\def\eeqno#1{\label{#1}\end{equation}}

\def\rarrow{\rightarrow }

\def\dleft{\rlap{{\it D}}\raise 8pt
\hbox{$\scriptscriptstyle\Leftarrow$}}
\def\dright{\rlap{{\it
D}}\raise 8pt\hbox{$\scriptscriptstyle\Rightarrow$}}

\def\cmss{{\rm cm~s^{-2}}}
\def\pc{~{\rm pc}}

\def\msun{M_{\odot}}
\def\az{a_{0}}
\def\azs{a_{0}^2}

\def\l0{\ell_{0}}

\def\rar{\rightarrow}
\def\s{\sigma}

\def\a{\alpha}
\def\b{\beta}
\def\c{\gamma}
\def\l{\lambda}

\def\f{\phi}
\def\t{\theta}

\def\k{\kappa}
\def\r{\rho}
\def\m{\mu}
\def\n{\nu}
\def\z{\zeta}

\def\vinf{V\_{\infty}}

\def\av#1{\langle#1\rangle}

\def\A{\mathcal{A}}

\def\F{\mathcal{F}}
\def\L{\mathcal{L}}

\def\SS{\mathcal{S}}

\def\o{\omega}

\def\vp{{\bf p}}

\def\a{\alpha}
\def\xlimin{{x\rarrow\infty \atop{\raise 1pt\hbox to 30pt
{\rightarrowfill}}}}
\def\limlim#1#2{{#1\rarrow #2 \atop{\raise 1pt\hbox to 30pt
{\rightarrowfill}}}}
\def\eps{\epsilon}
\def\vr{{\bf r}}

\def\vv{{\bf v}}

\def\vg{{\bf g}}

\def\va{{\bf a}}
\def\vA{{\bf A}}

\def\vK{{\bf K}}

\def\S{\Sigma}

\def\grad{\vec\nabla}
\def\div{\vec \nabla\cdot}
\def\gf{\grad\phi}

\def\L{\mathcal{L}}

\def\G{\mathcal{G}}
\def\fN{\f\_N}
\def\gfN{\gf\_N}
\def\vgN{\vg\_N}
\def\gN{g\_N}

\def\m{\mu}
\def\a{\alpha}
\def\b{\beta}

\def\n{\nu}

\def\F{\mathcal{F}}

\def\_#1{_{\scriptscriptstyle #1}}
\def\^#1{^{\scriptscriptstyle #1}}

\def\azg{\A_0}

\def\RM{r\_M}

\def\oot{\frac{1}{2}}

\def\tpg{2\pi G}
\def\fpg{4\pi G}
\def\epg{8\pi G}

\def\vrf{\varphi}

\def\gps{\grad\psi}
\def\gvrf{\grad\vrf}

\def\Sdz{\S^0\_D}
\def\Sbz{\S^0\_B}
\def\SB{\Sigma\_B}
\def\SD{\Sigma\_D}
\def\SP{\Sigma_p}
\def\SM{\Sigma\_M}
\def\Spz{\Sigma^0\_P}

\begin{document}
%\eqsec  % uncomment this line to get equations numbered by (sec.num)
\title{Central-surface-densities correlation in general MOND theories}

\author{Mordehai Milgrom}
\affiliation{Department of Particle Physics and Astrophysics, Weizmann Institute, Rehovot, Israel 7610001}

\begin{abstract}
It is shown that the foundational axioms of MOND alone predict a strong correlation between a bulk measure of the baryonic surface density, $\SB$, and the corresponding dynamical one, $\SD$, of an isolated, self-gravitating object, such as a galaxy. The correlation is encapsulated by its high- and low-$\SB$ behaviors.
For $\SB\gg\SM\equiv \az/\tpg$ ($\SM$ being the critical MOND surface density) one has $\SD\approx\SB$. More specifically, their difference -- which would be interpreted as the contribution of dark matter -- is $\SP=\SD-\SB\sim\SM\ll\SB$. In the deep-MOND limit, $\SB\ll\SM$, one has $\SD\sim (\SM\SB)^{1/2}$. This is a primary prediction of MOND, shared by all theories that embody its basic tenets.
Sharper correlations, even strict algebraic relations, $\SD(\SB)$, are predicted in specific MOND theories, for specific classes of mass distribution -- e.g., pure discs, or spherical systems -- and for specific definitions of the surface densities. I proceed to discuss such tighter correlations for the central surface densities of axisymmetric galactic systems, $\Sbz$ and $\Sdz$. Past work has demonstrated such relations for pure discs in the AQUAL and QUMOND theories. Here I consider them in broader classes of MOND theories.
For most observed systems, $\Sdz$ cannot be determined directly at present, but, in many cases, a good proxy for it is the acceleration integral $\G\equiv\int\_0\^\infty g_rd\ln~r$, where $g\_r$ is the radial acceleration along a reflection-symmetry plane of a system, such as a disc galaxy.
$\G$ can be determined directly from the rotation curve. I discuss the extent to which $\G$ is a good proxy for $\Sdz$, and how the relation between them depends on system geometry, from pure discs, through disc-plus-bulge ones, to quasispherical systems.
\end{abstract}
%\keywords{}
%\pacs{04.50.Kd, 98.80.Jk}
\maketitle
\section{Introduction}
The MOND program\footnote{MOND, which stands for ``modified Newtonian dynamics'', accounts for the mass anomalies in the Universe, not by invoking dark matter, but by an appropriate modification of Newtonian dynamics and general relativity \cite{milgrom83}. It is extensively reviewed in Refs. \cite{fm12,milgrom14,milgrom20,mcgaugh20,merritt20,bz22}.}  starts with the following basic tenets, which a MOND theory is required to obey:\footnote{We also often consider theories that go beyond these tenets; but as a start, we try to abide by them.}$^,$\footnote{MOND is defined by its basic tenets, from which follow a large number of primary predictions. One can write various theories that embody these basic axioms -- and hence share the primary predictions -- but which can differ on what they predict for secondary phenomena, or on details of the primary predictions.} 1. It introduces a new constant, $\az$, with the dimensions of acceleration. 2. A MOND theory should converge to standard dynamics in the formal limit $\az\rar 0$, namely, when all system attributes with the dimensions of acceleration become much larger than $\az$. 3. In the opposite, deep-MOND limit (hereafter DML) -- which is approached, formally, by taking $\az\rar\infty$, while $G\rar 0$, so that $\azg\equiv G\az$ remains fixed -- the theory should become spacetime scale invariant. (See  Refs. \cite{milgrom09a,milgrom14a} for details of this requirement and many of its consequences.)
One also requires that the theory does not involve dimensionless constants that differ much from unity. This implies, for example, that the transition from the DML to standard dynamics occurs not only around $\az$, but also within a range of order $\az$ around this value.
\par
The specific MOND theories studied to date, which embody the above tenets, were shown to predict a strong correlation between the characteristic surface density of a baryonic body and that of the putative -- but fictitious or ``phantom,'' according to MOND -- dark-matter halo.
For example,
for an axisymmetric galactic system -- such as a disc galaxy -- they predicts a strong correlation between the central surface density -- i.e., the column density along the symmetry axis -- of the baryonic body, $\Sbz$, and that of the phantom halo, $\Spz$.
\par
The essence of this central-surface-densities  correlation (hereafter CSDC) is captured by its two limits, according to whether $\Sbz$ is smaller or larger than the critical MOND surface density
\beq \SM\equiv \frac{\az}{\tpg}= 138(\az/1.2\times 10^{-8}\cmss)\msun\pc^{-2}. \eeqno{sigmam}
\par
For ``high-acceleration'' systems with $\Sbz\gg\SM$, it is predicted that $\Spz\sim\SM$. For systems in the DML, with $\Sbz\ll \SM$, it is predicted that
\beq  \Spz=\eta(\SM\Sbz)^{1/2},   \eeqno{yamyamu}
with $\eta=O(1)$.\footnote{It is often stated that because of the spacetime scale invariance of the DML, $G$ and $\az$ cannot appear separately in DML relations, only the product $\azg=G\az$; this appears not to be the case in Eq, (\ref{yamyamu}) since $\SM\propto\az/G$. Note, however, that for the statement to hold it is required that all system attributes in the relation are normalized such that they have scaling dimensions that match their $[t][\ell][m]$ dimensions. This is the case for $\Sbz$, but not for $\Spz$, which has dimensions of $m\ell^{-2}$, but scaling dimensions $-1$, not $-2$. If instead, we write the relation for $\S_p^{0*}\equiv G\Spz$, which has dimensions $\ell t^{-2}$, indeed only $\azg$ appears in the relation.} The exact interpolation between the two limits, and the exact value of $\eta$, depend on the specific MOND theory, and generally may depend also on the specific mass distribution.\footnote{Detailed predictions find $\eta$ between 2 for discs and $\approx 4$ for spheres.}
\par
The CSDC is sometimes plotted in terms of the total, ``dynamical'' central surface density, $\Sdz=\Sbz+\Spz$, which is predicted to converge to $\Sbz$ for $\Sbz\gg\SM$, and to $\Spz$ in the DML.
\par
As shown in Sec. \ref{rough}, the MOND basic tenets alone -- without resort to a specific theory -- predict such a correlation between the baryonic and dynamical bulk surface densities, such as the average one, or the surface density at the half-mass radius.
This makes such a correlation a primary MOND prediction.
\par
Instead, however, we concentrate here on the correlations between {\it central} surface densities of axisymmetric systems, as it emerges in specific MOND theories, for the following reasons: 1. The correlation between bulk surface densities, while it is more general, and not theory dependent in following from only the basic tenets, is not as tightly pinpointed as correlations between central surface densities in specific theories. For example, AQUAL \cite{bm84} and  QUMOND \cite{milgrom10}\footnote{The acronyms standing for ``aquadratic lagrangian'' and ``quasilinear MOND'', respectively.} predict for pure thin discs, an exact, algebraic relation (as opposed to a mere correlation) between $\Sbz$ and $\Spz$ (or $\Sdz$), which gives $\eta=2$ \cite{milgrom16}. 2. The basic tenets, dealing only with the two limits, are capable of informing us only on the asymptotic branches of a CSDC, while specific theories should predict its full extent.
3. $\Sdz$ is much easier to estimate from observations than bulk, dynamical surface densities -- using, e.g., rotation-curve data (see Sec. \ref{rcu}). 4. $\Sdz$ lends itself more easily to calculation in a given theory, because it requires the dynamical density only along the symmetry axis, not everywhere in the field.
\par
Past observational studies relevant to the issue \cite{ms05,spano08,donato09,lelli16} -- which I shall describe in more detail in Sec. \ref{done} -- confirm various aspects of the MOND prediction as elucidated in Refs. \cite{brada99,ms05,ms08,milgrom09c,milgrom16}.
\par
All the relevant predictions to date were derived either based on the naive algebraic relation between the MOND and Newtonian accelerations (from Ref. \cite{milgrom83}), or as they follow in the two workhorse, modified-gravity (hereafter MG) MOND theories,
AQUAL and  QUMOND.
\par
My main aim in the present paper is to consider the correlation as predicted in more general classes of MOND theories. In particular, I will discuss the value of the coefficient $\eta$ in the different theories.
\par
A secondary aim is to assess the procedure that we need to use when estimating $\Sdz$ from the observed rotation curve of a galaxy.
\par
In Sec. \ref{definitions}, I give the definitions of the various central surface densities, and  describe ways to determine them observationally, in particular, using rotation-curve data. I defer the discussion of some history of the subject to Sec. \ref{done}.
In Sec. \ref{rough}, I show how a correlation between baryonic and dynamical surface densities follows from the basic tenets of MOND.
In Sec. \ref{theories}, I describe the classes of more general modified-gravity MOND theories that are used in  Sec. \ref{predictions} to study the CSDC.
Section \ref{summary} is a summary.

\section{Definitions and procedures\label{definitions}}
In modified-gravity theories of nonrelativistic MOND, the acceleration of test particles in a gravitational field of a density $\r(\vr)$ is given by a gradient of the MOND potential, $\f(\vr)$,
\beq \ddot\vr=\vg=-\gf.   \eeqno{ia}
Unlike the Newtonian potential $\fN(\vr)$, which is the solution of the Poisson equation $\Delta\fN=\fpg\r$, and gives rise to the Newtonian acceleration field, $\vgN$, $\f$ solves some MOND equation that
incorporates the basic tenets of MOND.
\par
The predicted MOND acceleration field, $\vg$, should be identified with the {\it observed} acceleration field. So, one can defines a dynamical density distribution
\beq \r\_D\equiv -(\fpg)^{-1}\div\vg.  \eeqno{icha}
This auxiliary density would be required in Newtonian dynamics to reproduce the MOND gravitational field via the Poisson equation.
The difference,
\beq  \r\_p(\vr)=\r\_D(\vr)-\r(\vr),   \eeqno{iicha}
is called the phantom density \cite{milgrom86}, which in the context of a Newtonian interpretation of the observed (MOND) acceleration field has to be attributed to the putative dark matter.
\par
Some MOND results can be described, for convenience, in terms of this phantom density. Some examples of such a device can be found in Refs. \cite{milgrom86,ms08,oria21}.
\par
Here, we are interested in the ``central surface densities'' of the baryons, of the phantom mass, and the total, dynamical one, $\Sbz,~\Spz,~\Sdz$, respectively, of an isolated galactic system.\footnote{When the system is not isolated, in that it is falling in some external field, the internal dynamics is modified due to the MOND external-field effect.} This concept is relevant for an axisymmetric object -- such as an idealized disc galaxy, or an axisymmetric pressure-supported galaxy -- and is defined as the column density
of the respective component along the symmetry axis, taken as the $z$ axis. It is called so since it is the projected central surface density when the system is viewed along the axis.
We consider systems that are also reflection symmetric about a plane perpendicular to the $z$ axis, taken as the $x-y$ plane, which is many times a good approximation.
Then
\beq \Sbz=2\int_0^{\infty}\r^0\_B(z)dz=-\frac{1}{\tpg} \int_0^{\infty}\div\vgN(z)dz,~~~~\Sdz=2 \int_0^{\infty}\r^0\_D(z)dz=-\frac{1}{\tpg} \int_0^{\infty}\div\vg(z)dz, \eeqno{iiia}
 where superscript ``0'' on $\r$ indicates density on the $z$ axis.

\subsection{Determining the central surface densities  \label{determine}}
$\Sbz$ is directly measured -- at least in principle, and not always easily -- as the deprojected central surface density of the baryons, with contributions from stars and gas, etc.  $\Sdz$ includes the contribution of phantom matter (by MOND), or of dark matter (according to this paradigm). This cannot be detected (in MOND), and has not been detected (as dark matter); so, to measure $\Sdz$ we would need to map the acceleration field in some cylindrical volume around the symmetry axis, so $\div\vg$ can be determined there.
\par
In spherical systems, it is enough to measure the radial acceleration profile, since we then have from Eq. (\ref{iiia})
\beq \Sdz=\frac{1}{\tpg}\int_0^\infty\left(\frac{2g\_r}{r}+\frac{dg\_r}{dr}\right)~dr
=\frac{1}{\tpg}\int_0^\infty\frac{2g\_r}{r}dr=\frac{\k}{\tpg}\G, \eeqno{iva}
where
\beq  \G\equiv\int_0^\infty g\_r ~d{\rm ln}~r, \eeqno{maritew}
 and $\k=2$. {\it Here and below I take inward accelerations, $g\_r$, to be positive.}
The ``acceleration integral'', $\G$, a certain average of the radial acceleration, plays an important role in the present context.
\par
In deriving Eq. (\ref{iva}), I used $g\_r(\infty)=0$, and I also assumed that $g\_r(0)=0$, otherwise the integral diverges at least logarithmically at the origin.\footnote{The acceleration does not vanish at the origin if the density diverges at least as fast as $1/r$ there, in which case $\Sdz$ indeed diverges.}
\par
$\G$ can be measured with various degrees of accuracy for quasispherical systems, such as elliptical galaxies or galaxy clusters, using methods detailed in the literature, such as hot gas hydrostatics, or weak lensing. Such systems tend, however, to have $\Sbz\gg\SM$ and they would probe the ``less interesting'' region of high accelerations.
\subsubsection{Disc galaxies -- using rotation curves  \label{rcu}}
Our main concern here is with disc galaxies, because they offer the most extensive and most accurate sample for studying the CSDC.
As we shall see below, the predictions made by modified-gravity MOND theories concern the phantom and dynamical column densities along the symmetry axis.
However, except perhaps for the Milky Way,\footnote{The Milky Way also has $\Sbz\gg\SM$.} where we can, in principle, use test particles away from the plane to map the potential field (using, e.g., tidal streams; see, for example, Ref. \cite{ibata23}), we can only measure in a relatively robust way, at present, the acceleration field in the plane of the disc, using rotation curves: $g_r(r)=V^2(r)/r$.
\par
In the framework of the dark-matter paradigm, one can approximate the halo by a spherical model, determine its structural parameters (e.g., its characteristic density and size), and then calculate its $\Spz$. Or, one can determine $\Spz$ from Eq. (\ref{iva}) with $\k=2$, using the excess contribution to the acceleration integral, over that of the baryons.
\par
In MOND, however, the phantom body of a disc galaxy is made of a phantom disc and a phantom bulge \cite{milgrom01}. We thus need a relation similar to (\ref{iva}) for more general mass distributions with our assumed symmetry.
\par
Relation (\ref{iva}) for a spherical system can, in fact, be generalized to spheroids of arbitrary axes ratio, $q$, having a density distribution (in cylindrical coordinates)
\beq  \r(r,z)=\hat\r\left(r^2+\frac{z^2}{q^2}\right).   \eeqno{spheroid}
I show in Appendix \ref{appendix1} that If $\Sdz$ is the column density of such a spheroid along the symmetry axis, and $g_r(r)$ is the acceleration in the $x-y$, symmetry plane, then relation (\ref{iva}) holds with $\k=1+q$. For a thin planar disc ($q=0$), we have $\k=1$.
\par
Since any surface-density distribution in a thin disc can be written as a sum of infinitely flattened homoeoids (see, e.g., Sec. 2.6-1 of Ref. \cite{bt87}), relation (\ref{iva}) with $\k=1$ holds for any thin disc. This reproduces, as a special case, the relation found in Toomre \cite{toomre63} for thin discs.
\par
As another example of mass models with controllable degree of flatness,
for Miyamoto-Nagai models, having a potential field
\beq  \f(r,z)=-\frac{MG}{\{r^2+[a+(z^2+b^2)^{1/2}]^2\}^{1/2}},   \eeqno{marews}
I derive in Appendix \ref{appendix2}:
\beq \k=(1+\l)^2\int_0^1\frac{x^3(3+\l x)dx}{(1+\l x)^3\sqrt{1-x^2}},  \eeqno{nanmana}
where $\l=a/b$. As $\l$ goes from $0$ (Plummer sphere) to $\infty$ (Kuzmin disc), $\k$ goes, monotonically, from 2 to 1.
\par
Relations like Eq. (\ref{iva}) are based on the linear Poisson equation, and their two sides are additive.  It can thus be combined linearly for any number of aligned components with our symmetry; e.g., a disc-plus-sphere mass (whether real or phantom).
\par
To avoid possible confusion, note that when we discuss such relations in the context of MOND (which is nonlinear) they should always be understood to apply not to the baryonic surface density of the system, but to the dynamical density distribution predicted by MOND -- baryonic-plus-phantom. Hence, in the present context, such surface densities are designated with a subscript ``D.''
Thus, such relations are not additive with respect to the baryonic components of the mass distribution. Given a baryonic distribution, we first have to determine the MOND, dynamical distribution, identify different components of it, and then apply such relations.
\par
For a system made of several components with different $\k$ values, the acceleration integral, and the central surface density, are each the sum of contributions of the different components, $\G=\sum_i \G_i$, $\Sdz=\sum_i\S\_{Di}^0$. They are related as in Eq. (\ref{iva}), with an effective coefficient $\bar\k$, which, however, depends on the relative contributions of the components.
\beq \bar\k= \G^{-1}\sum_i\k_i \G_i=\Sdz\left[\sum_i\frac{\S\_{Di}^0}{\k_i}\right]^{-1}.  \eeqno{avava}
To determine $\bar\k$ from the first expression in Eq. (\ref{avava}) we would need to know the individual contributions to the full rotation curve. The second expression is more useful, as it requires only knowledge of the separate contributions to the central surface density, which is easier to estimate.
For example, a mass distribution that is a combination of oblate spheroids (with $0\le q_i\le 1$) has $1\le\bar\k\le 2$.
\par
When testing MOND predictions, the contributions of the different phantom components to $\Spz$, and hence to $\Sdz$, are predicted. So
if we can estimate the $\k$ values of the components, we can get a good approximation of $\bar\k$ from the second equality in Eq. (\ref{avava}).
For example, AQUAL and QUMOND predict that in the DML, where $\Sbz\ll\SM$, $\Sdz\approx\Spz$ picks up equal contributions from a phantom disc, for which $\k\approx 1$, and from an oblate spheroid, for which $\k$ should be between 1 and 2. Then, from Eq. (\ref{avava}), $\bar\kappa$ should be between 1 and 4/3.
For example, the exact value of $\bar\k$ predicted by AQUAL and QUMOND for a deep-MOND Kuzmin disc is $\bar\k=4/\pi$ \cite{milgrom16} (see also Sec. \ref{kuzrel} below).\footnote{For such a Kuzmin disc, AQUAL and QUMOND predict analytic expression for $\Sdz$ and $\G$; see Sec. \ref{aqumo} below.}
\par
Such accuracy in estimating $\bar\k$ is good enough for testing the MOND predictions, since, at present, $\G$ and $\Sbz$ themselves are not determined to better accuracy, and, because the span in $\Sbz$ studied in existing galaxy samples -- of several orders of magnitude -- is much larger than this uncertainty factor in the predicted $\Sbz/\G$.\footnote{It should be realized that such values of $\k\sim 1-2$ apply for most observed galaxy mass distributions of a disc plus an oblate-to-spherical bulge. But for general mass distributions (with our symmetry) there is no tight relation between $\Sdz$ and $\G$, and the justification of using $\k\sim 1$ has to be checked individually.
As extreme counterexamples, for a very long cylinder of constant density, the column density along the symmetry axis goes to infinity as the length of the cylinder does, but $\G$ remains finite; thus in this case $\k\rar\infty$. This is also true of a prolate spheroid with $q\gg 1$, for which $\k=1+q$. On the other hand, for a hollow, thin cylindrical shell, for which $\Sdz=0$, $\G$ is finite, and for infinite length becomes $\tpg \S_s$, where $\S_s$ is the surface density of the thin shell; so $\k=0$.}
\par
By employing Eq. (\ref{iva}), we turn a MOND prediction of a $\Sbz-\Spz$ or $\Sbz-\Sdz$ correlation to a $\Sbz-\G$ one, with some increase of the scatter due to the uncertainty in $\bar\k$. The great advantage of the latter is that it lends itself to testing, since both quantities in it can be measured, unlike $\Spz$ or $\Sdz$ which are very difficult to measure directly.
\par
Note that this prediction of the $\Sbz-\G$ relation does not follow from, and is independent of, the MOND prediction of rotation curves. The latter require a knowledge of the full baryonic mass distribution and predicts the full rotation curve. The former uses just one attribute of the mass distribution, $\Sbz$, and predicts from it one global attribute of the rotation curve.
We see already in the spherical case that two systems can have the  same  baryon central surface density, but different density distributions, and hence different rotation curves, but they are predicted to have approximately the same $\G$.
\par
Examples of other MOND predictions of global rotation-curve characteristics from global baryonic properties are: 1. The mass-asymptotic-speed relation, $MG\az=\vinf^4$, which predicts the constant asymptotic rotational speed, $\vinf$, of isolated bodies from their total baryonic mass, $M$ \cite{milgrom83}. 2. A prediction of the mass-weighted root-mean-square rotational speed of an isolated galaxy that is fully in the deep-MOND regime from its total mass: $(2/3)(MG\az)^{1/2}=2\pi M^{-1}\int_0^\infty r\S(r)V^2(r)dr$ \cite{milgrom12}.

\section{What has been done to date  \label{done}}
From the MOND vantage point, the first statement relevant to our discussion was made when it was pointed in Ref. \cite{brada99} that according to MOND, the gravitational acceleration produced by the fictitious dark matter cannot much exceed $\az$. This can be stated, alternatively, as ``according to MOND, the central surface density of the phantom halo cannot much exceed $\SM$''.
This follows from relations such as Eq.(\ref{iva}), which shows that the characteristic surface density in units of $\SM$ is some average over the acceleration in units of $\az$.
Reference \cite{brada99} based their derivation on either AQUAL, or the algebraic formulation, whereby $\m(g/\az)\vg=\vgN$.
\par
Induced by the prediction in Ref. \cite{brada99},  Ref. \cite{ms05} then used a sample of disc galaxies and fitted their rotation curves with baryons plus a cored-isothermal-sphere, dark-mater halo. It showed (in their Fig. 4) that the product of the ``halo'' central density and its core radius is bounded by, and concentrated strongly near a value of $\approx 10^2\msun\pc^{-2}\approx\SM$. This translates into the maximum acceleration of the model halo being somewhat below $\az$, in agreement with the prediction of Ref. \cite{brada99}.
The product itself is approximately the central surface density of the ``halo'' as defined in Eq. (\ref{iiia}). So the findings of Ref. \cite{ms05} are tantamount to the distribution of halo central surface densities of their galaxy sample, being narrowly peaked near $\approx\SM$. This sample was, however, lacking galaxies deep in the MOND limit.
\par
Other studies have found similar results, without noting that these vindicate a MOND prediction. For example, like Ref. \cite{ms05}, Ref. \cite{spano08} fitted rotation curves of a sample of disc galaxies including a cored-isothermal-sphere halo model, while Ref. \cite{donato09} fitted another set of rotation curves with Burkert model halos (which is also characterized by a central density and a core radius). Both found for their samples, a distribution of the ``halo'' central surface densities that is strongly peaked at a value of $\approx 140\msun\pc\approx\SM$.
\par
On the theoretical side, using the algebraic relation $\m(g/\az)\vg=\vgN$, Ref. \cite{ms08} calculated the phantom surface-density profiles -- in a plane perpendicular to the axisymmetry axis -- for a point mass, for two equal masses of different separations, and for a thin rod of various lengths [in units of their MOND radius, $\RM=(MG/\az)^{1/2}$]; all these for different choices of $\m(x)$.
The corresponding values of $\Spz$ for all these models can be read from the figures in Ref. \cite{ms08} (given there in units of $\az/G=2\pi\SM$).
\par
All the above results concern only the high-surface-density end of the CSDC, predicted to give $\Spz=\c\SM$, with $\c=O(1)$.
\par
In Ref. \cite{milgrom09c}, I considered the problem in some generality, using AQUAL, with the observational results of Refs. \cite{ms05,spano08,donato09} in mind. As noted above, those studies found a quasiuniversal value of $\Spz\approx\SM$ for the galaxy samples they employed, as MOND predicts for high-acceleration systems (see Sec. \ref{roughnewt}) below.
However, I noted that MOND does not predict a universal value of $\Spz$. This is the case only for high-surface-density systems, but for
low-surface-density ones, one predicts a correlation of the form $\Spz\propto (\SM\S_b)^{1/2}$, where $\S_b$ is some measure of the average surface density (not the central surface density, which I use in the present paper). The proportionality factor was estimated to be of the order of a few and to depend on the exact form of the interpolating function and on the exact mass distribution. The apparent quasiuniversality was attributed to the lack of truly low-surface-density galaxies in the sample studied. It was shown, with a few examples of such galaxies, that their values of $\Spz$ indeed fall much below $\SM$, and are consistent with the MOND predictions for the low-surface-density regime.
\par
By far the most extensive observational study of the correlation was described in Ref.  \cite{lelli16}  (see also Ref. \cite{swaters14} for a study with a much smaller sample, discussed in some detail in Ref. \cite{lelli16}). It shows the correlation between two proxies of the baryonic and dynamical central surface densities for 135 disc galaxies in the SPARC sample. The proxy for the baryonic central surface density was the stellar value, which may underestimate the baryonic value somewhat because of the contribution of the galactic gas.\footnote{In most cases gas is extended and, even when important globally, contributes little at the center; however, there are gas-dominated galaxies where the gas contribute materially to $\Sbz$.} The proxy for the dynamical central surface density was the acceleration integral, using Eq. (\ref{iva}), with $\k=1$ which, however, is appropriate only for pure discs.
\par
If one wishes to learn from the correlation about dark-matter halos, then a value of $\k$ nearer 2 would be more appropriate in the low-$\Sbz$ region where the spheroidal halo dominates. However, as we saw in Sec. \ref{rcu}, when testing the MOND prediction, a value of $\k$ between 1 and 4/3 is more in order; in this case the study of Ref. \cite{lelli16}, underestimates $\Sdz$ only a little.
\par
While Ref. \cite{lelli16} plotted a measure of $\Sdz$, MOND makes the finer prediction, that of $\Spz$, which is harder to separate from $\Sdz$ in the high-surface-density regime, because there $\Spz$ contributes very little to $\Sdz$. It is however, important to test this prediction separately from the grosser prediction of $\Sdz\approx\Sbz$.\footnote{In a related study, Ref. \cite{tian19} used the SPARC sample with fits for the separate contributions of the baryons and the phantom components to the rotation curves, to plot, separately, the phantom acceleration vs the baryonic one. They find, indeed that the phantom acceleration achieves a maximum at a value $\sim\az$ as predicted in Ref. \cite{brada99}. }
\par
Next came a derivation of a universal $\Sbz-\Sdz$ algebraic relation for pure baryonic discs predicted by AQUAL and QUMOND, which I describe next.
\subsection{Predictions of AQUAL and QUMOND for pure discs \label{aqumo}}
In Ref. \cite{milgrom16} I proved an exact result that holds in AQUAL and QUMOND, stating that the central dynamical surface density of a {\it pure disc} is a unique function of the baryonic central surface density:
\beq \Sdz=\SM \SS(\Sbz/\SM), ~~~~~~\SS(Y)=\int_0^Y\n(Y')dY',    \eeqno{vicha}
and $\n(Y)$ is the interpolating function of the respective theory.\footnote{The interpolating function $\m(X)$ that appears in AQUAL is represented here by $\n(Y)$ defined such that if $Y=X\m(X)$, then $X=Y\n(Y)$.}
The contributions to $\Sdz$ of the disc (baryonic plus phantom), $\SD\^{0,disc}$, and the bulge (all phantom), $\SD\^{0,sph}$,  are
\beq  \SD\^{0,disc}=\SM\n(Y)Y,~~~~~~~\SD\^{0,sph}=\SM\int_0^Y[\n(Y')-\n(Y)]dY',~~~~~Y=\Sbz/\SM.   \eeqno{bulasar}
$\n$ has the asymptotes  $\n(Y)\rar Y^{-1/2}$ for $Y\ll 1$, and $\n(Y)\rar 1$ for $Y\gg 1$.
\par
In the high-central-surface-density limit, $\Sbz\gg\SM$, the integral picks up most of the contribution from $y\gg 1$, so $\SS(y)\rar y$ in this limit, hence, $\Sdz\rar\Sbz$. In the opposite limit, the integrand is well approximated by $Y^{-1/2}$, and we have $\Sdz\rar 2(\SM\Sbz)^{1/2}$.
It is also quite interesting to consider the finer prediction for the separate contribution, $\Spz$, of the phantom halo.
Since $\Sbz=\SM\int_0^{\Sbz/\SM}dy$, we have
\beq \Spz=\Sdz-\Sbz=\SM\int_0^{\Sbz/\SM}[\n(Y)-1]dY.    \eeqno{kalage}
Solar-System constraints imply that $\n(Y)$ has to approach 1 fast as $Y\gg 1$. So the asymptotic value of $\Spz$
is $\SM\int_0^{\infty}[\n(Y)-1]dY$, which should converge fast. The integral is of order unity. For example, for the choice of $\n(Y)=Y^{-1/2}$ for $Y\le 1$ and $\n(Y)=1$ for $Y\ge 1$, we get $\Spz\rar 2\SM$.
\par
In the DML branch of the relation ($\Sbz\ll\SM$), we have $\Sdz=2(\SM\Sbz)^{1/2}\gg\Sbz$. We see from Eq. (\ref{bulasar}) that in this limit half of the contribution to $\Sdz$ comes from $\SD\^{0,disc}$, and half from $\SD\^{0,sph}$.
\par
Are these exact and universal (disc-independent) relations in AQUAL and QUMOND predicted in a wider range of MG theories?
It turns out that they are not quite, as we shall see below.
\section{Predictions of the basic tenets alone \label{rough}}
Now that we have the machinery in place, and before we consider more precise and detailed predictions of the CSDC, I show how the basic tenets of MOND predict a correlation between baryonic and dynamical surface densities. This makes the correlation in its rough form a primary prediction of MOND. Predictions of exact details of the correlation, such as the values of the numerical coefficients that appear, do depend, however, on the specific theory, on the details of the mass distribution, and on the concrete definition of the surface densities.
\par
The basic tenets of MOND inform us in detail only on the two limits of the theory, but not on the exact way in which the interpolation between them occurs in various theories, and for different phenomena. As regards the transition itself, because $\az$ is the only new dimensioned constant, and we assume that no dimensionless parameters appear that differ much from unity, the transition is predicted to occur around $\az$, in acceleration, and within a range of that order; viz. the asymptotic behaviors are already good approximations below $\az/q$ and above $q\az$, for $q\not\gg 1$.
As a result, we can hope to derive from the basic tenets alone only the asymptotic limits of a correlation, but be assured that these limits apply -- each in its own regime -- almost everywhere except for a region of order $\SM$ around $\SB=\SM$.
\par
For the arguments here to hold, we need to apply them to baryonic surface-density attributes, $\SB$, that are a good measure of the overall or average Newtonian acceleration of the system, such as the average surface density, because then $\SB/\SM$ is a good indicator of whether we are in the MOND or the Newtonian regime.
\subsection{Newtonian limit  \label{roughnewt}}
A system of mass $M$ and characteristic size $R$, with $\SB\sim M/\pi R^2\gg\SM$, is well contained within its MOND radius $\RM=(MG/\az)^{1/2}$; namely, $R\ll\RM$. Whichever way we measure accelerations, and whether we are dealing with MG or modified inertia (MI) theories, the accelerations within $\sim\RM$ are predicted to be Newtonian.\footnote{In MI theories we have to assume that the orbits of the test-particle probes do not take them far outside $\RM$.} Hence the phantom density picks up contributions only from $r\gtrsim\RM$. In this region we already have approximate spherical symmetry, and the average, or central, phantom surface density, $\SP\sim g(\RM)/\tpg\sim\SM$, as can be seen, e.g., from Eq. (\ref{iva}).

\subsection{Deep-MOND limit  \label{roughdml}}
Consider a density distribution $\r(\vr)$ such that the system is fully in the deep-MOND regime. Generate from it a two-parameter family of density distributions
\beq  \r\_\l\^\a(\vr)=\a\l^{-3}\r(\l^{-1}\vr), \eeqno{yabana}
whereby all sizes are stretched by a factor $\l$ and all masses are multiplied by a factor $\a$. Scale invariance of the DML, and dimensional arguments, imply \cite{milgrom09a,milgrom14} that all kinematic accelerations, $\va$ (which are the same as the MOND gravitational accelerations, $\vg$) that appear in solutions of the theory for $\r$, scale to $\a^{1/2}\l^{-1}\va$ and $\a^{1/2}\l^{-1}\vg$, respectively, in the solutions for $\r\_\l\^\a$.\footnote{To see this, note that the scale invariance of the DML implies that only the combination $\azg=G\az$ can appear in a DML expression for the MOND accelerations. A change of units must leave any valid expression valid. Change the units of mass such that the values of masses $m\rar\a m$, the length units such that $\ell\rar\l\ell$, and the units of time so that $t\rar \l\a^{-1/4}t$. Since the only dimensioned constant of the theory, $\azg$, remains unchanged under this scaling, the scaled solutions satisfy the DML equations with the correct value of $\azg$. The kinematic accelerations then indeed scale as $\a^{1/2}\l^{-1}$.}
Masses scale as $m\rar \a m$; dynamical densities scale as $\r\_D(\vr)\rar \a^{1/2}\l^{-2}\r\_D(\l^{-1}\vr)$ (not as $\a\l^{-3}$ like $\r\_\l\^\a$ itself). All dynamical surface densities scale as $\SD\rar\a^{1/2}\l^{-1}\SD$ (at the appropriately scaled positions).
On the other hand,
any baryonic surface density scales as $\SB\rar\a\l^{-2}\SB$. Thus, with any definition of some $\SD$ and $\SB$, the dimensionless, scale-invariant, and mass-scaling-invariant quantity (again, calculated at the scaled positions)
\beq  \eta(\l,\a)=\frac{\SD}{(\SM\SB)^{1/2}}    \eeqno{covaim}
is, in fact, independent of $\l$ or $\a$, as long as $\l$ is not as small, or $\a$ is not as large, as to bring the system out of the DML.
\par
To recapitulate this important result: {\it It follows from the basic tenets of MOND alone that in any given MOND theory that embody them, the DML value of $\eta$ is constant within any such two-parameter family of galaxies.} For example, all thin exponential (pure) disc galaxies in the DML, or all DML galaxies described by a Plummer model, etc. should have the same value of $\eta$.
\par
This fact holds for any choice of $\SB$ and $\SM$, even if they are unrelated, for example, the dynamical central surface density and the baryonic surface density at the radius
containing 90 percent of the projected mass. However, with arbitrary choices of the two surface densities, $\eta$ is not a useful ratio to consider, because it could be very large or very small and could depend very strongly on the family $\r\_\l\^\a$.
\par
A similar statement is correct if we replace $\SD$ in the nominator of $\eta$ by any measure of the MOND (measured) accelerations, such as $\G$ (or $\G/\tpg$ if we want the ratio to remain dimensionless) because they scale in the same way as $\SD$ under the scaling of Eq. (\ref{yabana}).
\par
To proceed with the argument, and produce a significant correlation between the two quantities, encompassing all density distributions, with $\eta$ of order unity, we want to choose the two surface densities so that they are physically related; i.e., the dynamical and baryonic analogs of each other. In particular, we choose their definition such that in the Newtonian limit $\SD$ goes to $\SB$ -- for instance, they can be the central surface densities, or the average surface densities, $\av{\SD}$ and $\av{\SB}$. In this way we can also bring to bear the second MOND tenet, that of the ``correspondence principle'' in the Newtonian limit.
\par
Given the generating density distribution $\r(\vr)$, there is a demarcation line in the $\l-\a$ plane, $\a\propto \l^2$, where $\SB=\SM$. Much above the line the corresponding system is Newtonian, and much below it, it is in the DML. In the former region, $\eta\rar (\SB/\SM)^{1/2}$, since we chose the definitions of $\SB$ and $\SD$ so that $\SD\rar\SB$ in this limit. And, as we saw, in the latter region $\eta$ is some constant (for the family). The essence of our argument is that the dimensionless $\eta$ cannot make a big jump when going from one side to the other. So, the constant, DML value cannot be very different from the extrapolation of the Newtonian $\eta$ values to the demarcation line. The latter extrapolate to $\eta=1$ on the dividing line\footnote{In specific examples that we have, the actual value of $\eta$ on the dividing line is, in fact, somewhat larger than one, and the DML values of $\eta$ vary roughly between 2 for discs and $\sim 4$ for spheres.}; so the DML constant value must be $\eta=O(1)$. This follows from the tenet saying that the two limiting expressions of $\eta$ hold approximately to ``factor of order unity,'' down to the transition.
\par
Since this has to hold for all choices of $\r(\vr)$, we have, generally,
$\eta=O(1)$ in the DML for all bounded mass distributions.
\par
This establishes a general CSDC. Despite the impression that such a predicted CSDC can be rather loose, with large scatter, one has to note that the span of $\SB$ values to which we apply it in practice can be several orders of magnitude; so the predicted correlation would still be quite strong.
\par
The arguments here and in Sec. \ref{roughnewt} assume that the ratio of the baryonic surface density we employ to $\SM$ is a good measure of whether we are in the DML or the Newtonian regime. This would be the case for global measures of the surface density, such as $\av{\SB}$, or the surface density at the half-mass radius.
Beyond these general arguments that use essentially only the basic tenets, but predict a correlation with possibly large scatter, we would like, if possible, to get more refined predictions, at a cost of being, perhaps, less general, in being theory dependent, and dependent on the exact definitions of the surface densities involved.
This leads us to consider the central surface densities in the correlation, because, as I said above, $\Sdz$ is both easier to measure and to calculate in a given theory.
\par
In many disc galaxies, $\Sbz$ is close to a global measure of the surface density (e.g., for exponential discs), and for these galaxies the above arguments apply to the central surface densities, as well.
However, in principle, $\Sbz$ may be quite different from the average surface density of the galaxy, and hence its ratio to $\SM$ is not a good MOND-vs-Newtonian-dynamics indicator.
For example, consider a disc galaxy with a concentric, circular hole around its center, whereby $\Sbz=0$. This remains so all along the above $\l,~\a$ sequence; so the above argument does not work in this case.
Nonetheless, the exact relation (\ref{vicha}) holds for this case in AQUAL and QUMOND, and it dictates that $\Sdz=0$, as well, for all values of $\l$ and $\a$.\footnote{In this case, there is a region of negative phantom density along the $z$-axis \cite{milgrom86}, which results in $\Spz$ vanishing. There is also a region around the center where the acceleration integral, $\G$, picks negative contributions (the radial accelerations point outward). The Newtonian rotation curve would then give $\G=0$. But the observed (MOND) rotation curve will generally give $\G\not = 0$. $\G$ is then not useful for determining $\Sdz$, since in this case the effective $\bar\k=0$ in Eq. (\ref{avava}) is very different from unity.} This is a result that we could not have derived based on the above general argument, but requires reference to the specific theory.
\par
With these caveats and subtleties in mind, I now proceed to discuss the predictions of more concrete theories.

\section{Generalized theories  \label{theories}}
In this section, I describe the classes and subclasses of theories that I will use to generate predictions of a CSDC.
I shall consider in some detail only MG theories, with some comments on the not-less-promising class of MI theories. The reason for this preference it twofold. First, we do have concrete, full-fledged, and self-consistent examples of MG MOND theories, while we are still lacking such well-developed MI theories.
Second, as I explain in Sec. \ref{mit} below, the concept of phantom density, and hence of a dynamical surface density, is not well defined in MI theories.
\par
Nevertheless, we do have some general results for a certain class of MI theories, which I discuss briefly in Sec. \ref{mit}
\subsection{General class of modified-gravity MOND theories \label{genmg}}
The class of MG, MOND theories -- more general than AQUAL and QUMOND -- that I consider here in some more detail are governed by a Lagrangian density
\beq \L=\L_G  +\r(\oot\vv^2-\phi),  \eeqno{lagrangen}
whose gravitational part is
\beq \L_G=-\frac{\azs}{\fpg}F\left(\frac{\gf}{\az},\frac{\grad\psi\_1}{\az},\frac{\grad\psi\_2}{\az},...\right) .  \eeqno{lagra}
(Note that only the first derivatives of the potentials appear.)
They are a subclass of the more general theories described in detail in Ref. \cite{milgrom14b}.
\par
Of the scalar potentials $\f,~\psi_1,~\psi_2,...$, which are the gravitational degrees of freedom, only $\f$ couples directly to the matter density. Variation over the matter degrees of freedom gives the equation of motion $\ddot\vr=-\gf$, identifying $\f$ as the MOND potential, which dictates accelerations of test particles.
$F$ is a dimensionless function that depends on the potentials through the scalar products, $\gf\cdot\grad\psi_i$, and $\grad\psi_i\cdot\grad\psi_j$, for $j\ge i$.
\par
The field equations are
\beq  \azs\div\left(\frac{\partial F}{\partial\gf}\right)=\fpg\r=-\div\vgN,~~~~~~  \div\left(\frac{\partial F}{\partial\grad\psi\_i} \right)=0.\eeqno{eqfa}
\par
A useful property of this general class of theories, which I employ below, is that for one-dimensional systems, with spherical, cylindrical, or plane symmetry, the MOND acceleration, $\vg$, is a universal function of the Newtonian acceleration, $\vgN$ -- a function unique to the theory, but independent of the specific one-dimensional symmetry, or the specific mass distribution. We thus have
\beq  \vg=\n(\gN/\az)\vgN,~~~~~\vgN=-\frac{GM(r)\vr}{r\^3}.  \eeqno{onedim}
To see this, apply Gauss's theorem to the field equations in Eq. (\ref{eqfa}) for constant-$r$ surfaces, where $r$ is the coordinate that underlies the one-dimensional symmetry. All the acceleration-vector arguments of $F$ are collinear and have only an $r$ component; so they can be replaced by  $\gf\rar g=|d\f/dr|$,
$\grad\psi\_i\rar g\_i=|d\psi\_i/dr|$, and all scalar products can be replaced by products of the $g$s. $F$ then becomes a function of $\chi\equiv g/\az$ and $\chi_i\equiv g\_i/\az$, $F=\hat F(\chi,\chi_1,\chi_2,...)$, and the field equations become algebraic equations between $\chi,~\chi_i$
\beq \frac{\partial \hat F}{\partial\chi}(\chi,\chi_1,\chi_2,...)=\frac{\gN}{\az}, \eeqno{eqfarad}
\beq  \frac{\partial \hat F}{\partial \chi_i} (\chi,\chi_1,\chi_2,...)=0. \eeqno{eqpsarad}
\par
Equations (\ref{eqpsarad}) are used to solve for all the $g\_i/\az$ as functions of $g/\az$, substituted in Eq. (\ref{eqfarad}), which is then inverted to give Eq. (\ref{onedim}). The possibility to make all these inversions is a necessary requirement from $F$, for the theory to predict gravitational fields uniquely.\footnote{In AQUAL, for example, this is the requirement of ellipticity of the field equations, which requires $x\m(x)$ to be monotonic.}
\par
Equation (\ref{onedim}) defines the ``one-dimensional interpolating function'', $\n(Y)$.
\par
After some general results that follow for all the theories in the class, I shall concentrate, for more detailed calculations, on a yet smaller subclass -- the tripotential MOND (TRIMOND) theories discussed in Ref. \cite{milgrom23}.
While TRIMOND theories -- with AQUAL and QUMOND as special cases -- form a subclass, with a Lagrangian density of the type (\ref{lagra}), the generalizations of QUMOND presented in Ref. \cite{milgrom23a} are examples of MG theories that are not of this type, because their Lagrangians depend also on higher derivatives of the auxiliary potentials. These latter theories do not predict the algebraic relation (\ref{onedim}) for one-dimensional systems.
\par
Finally, as shown in Ref. \cite{milgrom14b}, all the theories in this class predict the DML virial relation for self-gravitating, isolated system of (pointlike) masses $m_p$,
\beq \sum_p \vr_p\cdot\vK_p=-(2/3)(G\az)^{1/2}[(\sum_p m_p)^{3/2}-\sum_p m_p^{3/2}],  \eeqno{galama}
where, $\vr_p$ are the positions of the masses, and $\vK_p$ are the forces they are subject to.
A special case is the deep-MOND two-body force for arbitrary masses,
 \beq {\rm K}(m_1,m_2,\ell)=\frac{2}{3}\frac{(\az G)^{1/2}}{\ell}[(m_1+m_2)^{3/2}-m_1^{3/2}-m_2^{3/2}] \eeqno{shasa}
($\ell$ is the distance between the masses).
Another important corollary is the DML mass-velocity-dispersion relation
\beq \s^2=\frac{2}{3}(MG\az)^{1/2}[1-\sum_p (m_p/M)^{3/2}],  \eeqno{shisa}
where $\s^2=M^{-1}\sum_p m_p\vv_p^2$ (velocities are in the rest frame of the center of mass), and $M=\sum_p m_p$ is the total mass. This reduces to $\s^4=\frac{4}{9}MG\az$ for systems made of many masses, with $m_p\ll M$.
\par
In particular, all these theories predict the same value of the $Q$ parameter, defined in Ref. \cite{milgrom12}, which was proposed as a possible discriminator between MOND theories.
\par
After considering in Sec. \ref{predictions} some general prediction of this class of theories, I derive more specific ones for the TRIMOND subclass, which I now recap briefly.
\subsection{TRIMOND recap  \label{trimond}}
Following is a summary of the aspects and equations of TRIMOND from
Ref. \cite{milgrom23} that we shall be needing here.
\par
The TRIMOND Lagrangian density is
\beq \L=-\frac{1}{\epg}[2\gf\cdot\grad\psi-\azs\F(x,y,z)] +\r(\oot\vv^2-\phi),  \eeqno{lagran1}
where
\beq x\equiv(\gps)^2/\azs,~~~y\equiv(\gvrf)^2/\azs,~~~z=2\gps\cdot\gvrf/\azs, \eeqno{xyz}
and $\F$ is a dimensionless function satisfying the basic tenets of MOND.
\par
This is a special case of the Lagrangian density (\ref{lagra}),
 special in two important regards:
(a) It involves only three gravitational potentials.
(b) The MOND potential, $\f$, couples directly to only one of the auxiliary potentials, $\psi$, and that linearly. This results in $\psi$ being the Newtonian field, and is straightforwardly solved for.
As a result, unlike the general case, TRIMOND theories do not require solving coupled equations.
\par
Acceleration of test bodies is given by
 \beq \ddot\vr_i=-\gf(\vr_i),   \eeqno{iitr}
where $\f$ is the MOND gravitational potential. The first of Eq. (\ref{eqfa}) becomes here
\beq \Delta\psi=\fpg\r, \eeqno{iiatr}
 and the two additional field equations are
\beq \div(\F_y\gvrf)+\div(\F_z\gps)=0,   \eeqno{vvvf}
\beq \Delta\f=\div(\F_x\gps)+\div(\F_z\gvrf)\equiv\fpg\r\_D,   \eeqno{mail}
where $\r\_D$ is the dynamical density (baryonic plus phantom).
\par
$\psi$ thus equals the Newtonian potential, and is solved for from Eq. (\ref{iiatr}). It is then substituted in Eq. (\ref{vvvf}), which becomes a nonlinear, second order equation in $\vrf$. Solving for it, and substituting both in the right-hand side of Eq. (\ref{mail}), we get a Poisson equation for $\f$.
\par
AQUAL is a special case with $\F_x=0,~\F_z=\eps$, a constant.
QUMOND is a special case gotten for the choice $\F_y=\F_z=0$.
\par
As a subclass of the above class of MG theories, for systems with one-dimensional symmetry, TRIMOND theories predict the algebraic relation (\ref{onedim}) between the MOND and Newtonian accelerations.
In such cases, the field equations (\ref{vvvf}) and (\ref{mail}) can be stripped of the divergences, and become algebraic (putting, here, $\az=1$),
\beq  \F_y(\gN\^2,g\_\vrf\^2,2\gN g\_\vrf)g\_\vrf+\F_z(\gN\^2,g\_\vrf\^2,2\gN g\_\vrf)\gN=0,   \eeqno{mumuiop}
from which $g\_\vrf$ can be solved for as a function of $\gN$. $\F$ has to be such that the solution for $g\_\vrf$ exists and is unique; the ellipticity of Eq. (\ref{vvvf}) should ensure this.
Equation (\ref{mail}) gives, in the same vein,
\beq  g=\F_x(\gN\^2,g\_\vrf\^2,2\gN g\_\vrf)\gN+\F_z(\gN\^2,g\_\vrf\^2,2\gN g\_\vrf)g\_\vrf.    \eeqno{mumulig}
Substituting the $g\_\vrf$ values gotten from Eq. (\ref{mumuiop}), we then get an equation of the form (\ref{onedim}).
\par
The Newtonian limit for $\az\rar 0$ requires that in this limit $\F_z\rar \eps$, $\F_y\rar\o$, and $\F_x\rar\b$, all constants, with
\beq \b-\eps^2/\o=1.   \eeqno{newman}
\par
Scale invariance in the DML (discussed in detail in Ref. \cite{milgrom23}) requires that there is a scaling dimension, $\a$, of $\vrf$, such that in the DML $\F$ goes to some $\F\_D(x,y,z)$ that has the property
\beq \F\_D(\lambda^{-4}x,\lambda^{2\a-2}y,\lambda^{\a-3}z)=\lambda^{-3}\F\_D(x,y,z).   \eeqno{siff}
It is then necessary and sufficient that the DML of $\F$ is of the form
\beq \F\_D(x,y,z)=x^{3/4}\F\_D[1,yx^{(\a-1)/2},zx^{(\a-3)/4}]\equiv x^{3/4}\bar\F\_D[yx^{(\a-1)/2},zx^{(\a-3)/4}],   \eeqno{niuter}
Namely, the DML of TRIMOND requires one to specify a function of two variables, compared with no such freedom in the DML of AQUAL and QUMOND (see Sec. \ref{fzcon} for an explicit example).

\subsection{Modified-inertia theories  \label{mit}}
MI MOND theories are defined in the following restricted sense \cite{milgrom94,milgrom99,milgrom22a,milgrom23b}: The force field (gravitational in our applications) is assumed to be unmodified, and is thus Newtonian, with the potential $\fN$. However, the equation of motion of a test mass in this field is not the Newtonian $\ddot\vr=\vgN=-\gfN$, but has the form
\beq \vA[{\vr(t)}, \az]=-\gfN  .\eeqno{gurata}
where $\vA$ -- having the dimensions of acceleration -- is a functional of the body's trajectory ${\vr(t)}$ [as opposed to a function of the local value of $\vr(t)$]. Reference \cite{milgrom23b} is a review of such theories.
\par
In such theories, {\it a MOND acceleration field is not defined}. Different particles can have different accelerations ($\ddot\vr$) at the same position in the field, depending on details of their trajectories, as explained in detail in Ref. \cite{milgrom23b}.\footnote{This is similar to what happens, e.g., in special relativity, where different electrons have different accelerations at the same position in an electric field, depending on their velocity vector.}$^,$\footnote{When there is  a conserved momentum, $\vp$, in such a theory, $\dot\vp$ is the same for all particles at the same position, and does define a field; but this quantity is not proportional to $\ddot\vr$, which is what we measure directly.}
{\it Thus the notions of dynamical density and phantom density do not exist, generally, and there is no well-defined sense in speaking of the phantom central surface density.}
\par
However, in practice, any procedure that results in some dynamical surface density, whether direct or indirect (e.g., using the acceleration integral $\G$ as a proxy), is based on some measurements of accelerations. As explained in Sec. \ref{rough}, the general predictions of the basic tenets of MOND then apply to MI theories as well.
\par
Even beyond such a general prediction, we may be able to make more specific predictions in some specific situations.
A prominent case in point is the use of rotation curves to determine the dynamical and phantom central surface densities using the procedure described in Sec. \ref{definitions}. This procedure uses the measured accelerations on exactly circular orbits, by measuring the radius $r$ and orbital velocity $V(r)$ of these trajectories.
Now, for such trajectories, the acceleration functional $\vA[{\vr(t)}, \az]$  in Eq. (\ref{gurata}) must, in fact, be a {\it function} of $r$ and $V$, and so, on dimensional grounds, must be of the form
\beq \vA[{\vr(t)}, \az]=\frac{V^2}{r}\m\left(\frac{V^2}{r\az}\right), \eeqno{mushiq}
where $\m(x)$ is universal for the theory.\footnote{In Ref. \cite{milgrom94}, I showed how $\m(x)$ is derived from the action underlying the theory, when there is one.}
Thus, for circular orbits\footnote{In practice, the test particles in discs of galaxies have orbits that depart from exact circular, such as by motions perpendicular to the disc. This causes the predicted velocities to be somewhat smaller than given by Eq. (\ref{mushiq}). This is particularly important in the inner parts. See discussion of this in Ref. \cite{milgrom22a}.} in an axisymmetric body, such modified-inertia theories predict an algebraic relation between the MOND centrifugal acceleration, $g_c$, and the Newtonian, gravitational acceleration, $\gN$, $g_c\m(g_c/\az)=\gN$, which can be inverted to read
\beq  g_c=\gN\n(\gN/\az),   \eeqno{mispher}
identical with Eq. (\ref{onedim}), which is valid for spherical systems in all the modified-gravity theories considered above.
\par
Thus, for such spherical systems, the modified-inertia, and the class of modified-gravity theories predict the same distribution of dynamical density [for the same choice of $\n(Y)$], provided we use rotation-curve data for the determination.
\par
For example, for the deep-MOND polytropes discussed in Ref. \cite{milgrom21}, the modified-inertia theories predict the same values of the coefficient $\eta$ in Eq. (\ref{yamyamu}) as the modified-gravity theories.
\par
For disc galaxies, relation (\ref{mispher}) still holds for exact circular orbits in the galactic plane, but it does not hold for MG theories. Nevertheless, numerical calculations tell us that  AQUAL and QUMOND predict rotation curves for typical galaxy models, such as an exponential disc, that do not differ much from those gotten from this relation \cite{brada95,chae22}. This leads us to expect that for such models, the $\G$ integral predicted by MG and MI for the same pure-disc distribution would not differ by more than a factor $\sim2$ (when a bulge is present, the difference is even smaller). So, if one uses $\G$ as proxy for $\Sdz$, then AQUAL and QUMOND and MI would predict similar $\Sdz$ up to a factor $\sim 2$ in the DML (nearer the Newtonian regime the differences are smaller).\footnote{We are speaking here of MI-MG similarity in predicting the $\Sbz-\G$ correlation, not the $\Sbz-\Sdz$ correlation, which AQUAL and QUMOND predict exactly.}
\par
For example, Kuzmin discs, which I discuss in more detail in Sec. \ref{kuzminspired}, lend themselves to analytic derivation of the rotation curve in MI and in the class of theories described in Sec. \ref{genmg}. The latter give results that coincide with those of AQUAL and QUMOND for a Kuzmin disc, for which we have in the DML \cite{brada95},
\beq g\_{MG}=(V\_\infty^2/h)u/(1+u^2),  \eeqno{kuzaza}
where $u=r/h$, and $h$ is the scale length of the disc. While
\beq g\_{MI}=(V\_\infty^2/h)u^{1/2}/(1+u^2)^{3/4}.   \eeqno{kuzami}
These give
$\G\_{MG}=(\pi/2)(V\_\infty^2/h)$, while $\G\_{MI}=(V\_\infty^2/h)\int_0^{\pi/2}(\sin\t)^{-1/2}d\t=2.62(V\_\infty^2/h)$, some 70 percent higher.

\section{Predictions of the correlation \label{predictions}}
Numerical calculations would be required to derive the predictions concerning the CSDC for general disc-plus-bulge systems in the general theories of Sec. \ref{genmg}. Here, I give some analytic results for this class of theories for spherical systems, and for a class of disc-plus-bulge mass models built on the same principle as the Kuzmin disc. I also discuss some results for general pure-disc systems.
\subsection{Spherical systems}
Consider first spherical systems. The algebraic relation (\ref{onedim}) between the MOND and Newtonian accelerations then holds in the general class of MG theories described above, and we
have from Eq. (\ref{iva})
\beq  \frac{\Sbz}{\SM}=2\int_0^\infty Y(r)d{\rm ln}~r,~~~~~\frac{\Sdz}{\SM}=2\int_0^\infty \n[Y(r)]Y(r)d{\rm ln}~r,~~~~~\frac{\Spz}{\SM}=2\int_0^\infty \{\n[Y(r)]-1\}Y(r)d\ln{r}, \eeqno{matews}
where $Y(r)=\gN(r)/\az$.
\par
Another useful expression for $\Sdz$ in terms of the baryonic density distribution is
\beq\Sdz=-\frac{1}{\tpg}\int_0^\infty\div\vg~dr=-\frac{1}{\tpg}\int_0^\infty\div[\n(\gN/\az)\vgN]~dr=2\int_0^\infty \n[Y(r)] \r_s(r)dr-\SM\int_0^\infty(\vgN/\az)\cdot\grad\n(\gN/\az)~dr.   \eeqno{farepo}
The second integral can be shown to vanish [with our assumption that $\gN(0)=0$], leaving us with
\beq\Sdz=2\int_0^\infty \n[Y(r)]\r_s(r)dr.  \eeqno{manishma}
\par
In the high-acceleration limit, $\Sbz\gg\SM$, contribution to the last integral in Eq. (\ref{matews}) comes from outside the mass, so we can put $Y(r)=MG/\az r^2$, where $M$ is the total mass. Changing variables to $Y$ and noting that $d\ln{r}=-d\ln{Y}/2$, we have in the limit
\beq  \frac{\Spz}{\SM}=\int_0^\infty [\n(Y)-1]dY~=O(1), \eeqno{matesa}
the same expression we found below Eq. (\ref{kalage}) for pure discs in AQUAL/QUMOND.
\par
{\it  This result is universal -- i.e., holds for any bounded mass -- for all the theories in the class,} since the phantom density picks up contribution only far from the mass, where we can take the field as spherical.
\par
In the opposite limit, $\Sbz\ll\SM$, we have $y\ll 1$ at all radii; so, in the integral in the second of Eqs. (\ref{matews}) we can put  $\n(Y)=  Y^{-1/2}$. We then have
$\Sdz\rar\eta(\SM\Sbz)^{1/2}$,  with the coefficient
\beq  \eta=\sqrt{2}\frac{\int_0^\infty Y^{1/2}(r)d{\rm ln}~r}{\left[\int_0^\infty Y(r)d{\rm ln}~r\right]^{1/2}}=\sqrt{2}\frac{\int_0^\infty M^{1/2}(r)r^{-2}dr}{\left[\int_0^\infty M(r)r^{-3}dr\right]^{1/2}}.   \eeqno{liouter}
\par
We see that $\eta$ depends somewhat on the exact mass distribution. It is independent, however, of the normalization and of the overall length scale of the mass distribution, both of which drop out from expression (\ref{liouter}), as long as they do not take us outside the DML. This follows, generally, from basic MOND tenets, as we saw in Sec. \ref{roughdml}.
\par
For spherical systems, $\eta$ is generally of order of a few. For a homogeneous sphere $\eta=2\sqrt{3}\approx 3.46$.
In Ref. \cite{milgrom21}, I calculated $\eta$ for DML polytropes. Fig. 11 there shows that, for isotropic-velocity-dispersion polytropes of all orders (which include isothermal spheres as a limiting case), $\eta$ varies only by a little:  $2\sqrt{3}\le\eta\le3.81$.\footnote{The largest value of $\eta=\sqrt{8\pi/\sqrt{3}}\approx3.81$ is calculated for a DML isothermal sphere.} The values of $\eta$ are also shown for anisotropic DML polytropes, for different values of the anisotropy parameter (Fig. 16 there). For Plummer spheres one has $\eta\approx 3.71$.
\par
We see that spheres have $\eta$ values $~3-5$, while discs tend to have $\eta\approx 2$.

\subsection{Kuzmin-inspired disc-plus-bulge systems  \label{kuzminspired}}
I shall discuss the CSDC for pure discs in Sec. \ref{puredisc} below. Disc-plus-bulge systems are more difficult to treat analytically.
\par
We can, however, derive some exact results in the class of theories described in Sec. \ref{genmg} for a family of disc-plus-bulge mass distributions. Such exact results are important anchors for some of our general discussion, even if they are derived only for toy examples.
The results for these mass models are the same in all the theories in the class, and are expressed only in terms of their one-dimensional interpolating function $\n(Y)$.
\par
These models are constructed based on an idea that generalizes the construction of the Kuzmin disc, as described in Ref. \cite{brada95}.
\par
I use these galaxy models here to test how well the acceleration integral $\G$ actually measures the dynamical central surface density $\Sdz$ and to examine the CSDC that our MG theories predict for them.
\par
To construct the baryonic models, start with a spherical density distribution $\r_s(R)$ with accumulated mass $M(R)$.
For the models, the Newtonian potential above the $x-y$, symmetry plane is that of our spherical mass centered on the $z$ axis at $z=-h$. The potential below the symmetry plane is a reflection of the one above. The Newtonian potential satisfies the Poisson equation
outside the symmetry plane, with the (baryonic) density
\beq \r\_B(r,z)=\r_s(R), ~~~~~~R=[r^2+(z+h)^2]^{1/2}  \eeqno{mammm}
 as a source (using cylindrical coordinates $r,~z$); this density distribution constitutes our model baryonic bulge. The Newtonian acceleration field outside the disc is
\beq  \vgN=-\frac{GM(R)}{R^3}(r,z+h).  \eeqno{hatyo}
The $z$ component of $\vgN$ just above the disc is
\beq g_{Nz}^+(r)=-\frac{GM(\bar R)h}{\bar R^3}, ~~~~~\bar R=(r^2+h^2)^{1/2},\eeqno{jalater}
and just below the disc $g_{Nz}^-(r)=-g_{Nz}^+(r)$. This jump implies that, for the Poisson equation to be satisfied everywhere, the mass distribution has to be supplemented with a thin disc of (baryonic in our context) surface density
\beq   \SB\^{disc}(r) =-\frac{1}{\tpg}g_{Nz}^+(r)=\frac{hM(\bar R)}{2\pi \bar R^3}.   \eeqno{mixuta}
If $\r_s(R)$ is nonincreasing, then so is $\SB\^{disc}(r)$.
The resulting mass distribution is a disc, and a bulge made of the two caps of the mass distribution $M(R)$.
Since we want the bulge mass to be finite, $\SB\^{disc}(r)$ goes asymptotically to the Kuzmin surface density $\SB\^{disc}(r)\propto \bar R^{-3}$.
For example, for a sphere of radius $R_0$ and constant density $\r_0$, we get a disc of constant surface density $2h\r_0/3$ up to $r=(R_0^2-h^2)^{1/2}$, and a Kuzmin disc beyond this, plus a constant-density lenslike bulge.\footnote{If we start with the center of the sphere above the plane, we end up with an hourglass-shaped bulge.} The Kuzmin disc is gotten when the whole spherical mass is below the plane; namely, when $M(R>h)$ is constant (e.g., for  a point mass).
\par
Despite the complexity of the MG theories described in Sec. \ref{genmg} -- they yield a system of many nonlinear, coupled equations -- and their generality, the MOND field equations in Eq. (\ref{eqfa}) can be solved analytically for the above mass models.
\par
The mass distribution and the Newtonian field above the plane are spherical with respect to the center at $r=0,~z=-h$; so, as shown in Sec. \ref{genmg}, there is a solution of the full MOND equations -- the one-dimensional solution -- whereby $\gf$ and the $\grad\psi_i$ are all aligned with $\gfN$. This solves the MOND equations above the symmetry plane with $\r\_B$ from Eq. (\ref{mammm}) as source, and its reflection solves them below the plane. All the vector fields being radial from some point, they have a vanishing curl. Thus, this MOND solution solves the field equations (\ref{eqfa}) with the divergences stripped. Namely,
\beq  \azs\frac{\partial F}{\partial\gf}=\gfN,~~~~~~  \frac{\partial F}{\partial\grad\psi\_i} =0,\eeqno{eqfastr}
which hold above and below the plane. But this means that this solution also satisfies all the jump conditions across the disc, dictated by $\SB\^{disc}$ from Eq. (\ref{mixuta}),\footnote{The jump conditions from Eq. (\ref{eqfa}) are that the perpendicular component of $\partial F/\partial\grad\psi\_i$ has to be continuous, and that of $\azs(\partial F/\partial\gf)$ has to be $\tpg\SB\^{disc}$.} since the Newtonian solution does. This establishes it as the MOND solution of the problem.\footnote{For this solution $\grad\psi_i=\n_i(\gfN|/\az)\gfN$, and in particular, $\gf=\n(\gfN|/\az)\gfN$, with $\n(Y)$ the one-dimensional interpolating function of the theory.}
\par
In particular, the MOND acceleration field above the disc is
\beq  \vg=-\n\left[\frac{GM(R)}{\az R^2}\right]\frac{GM(R)}{R^3}(r,z+h),  \eeqno{hatmush}
 with $\n(Y)$ the one-dimensional interpolating function of the theory.
The parallel component of $\vg$ is continuous across the disc. The jump in the perpendicular MOND field is interpreted as a dynamical surface density (baryonic+phantom)
\beq  \SD\^{disc}(r)=\n\left[\frac{GM(\bar R)}{\az \bar R^2}\right]\SB\^{disc}(r).  \eeqno{ferush}
The baryonic central surface density is
\beq \Sbz=2\int_h^\infty\r_s(\bar z)~d\bar z+\frac{M(h)}{2\pi h^2}=\frac{1}{\pi}\int_h^\infty \frac{M(\bar z)}{\bar z^3}d\bar z,   \eeqno{bzaza}
where the second equality is gotten by writing $\r_s(\bar z)=M'(\bar z)/4\pi \bar z^2$, and integrating by parts.
\par
The dynamical (MOND) central surface density picks up a contribution, $\SD\^{0,disc}$, from the disc (baryonic plus phantom), and one, $\SD\^{0,sph}$, from the bulge, with
\beq \SD\^{0,disc}= \n\left[\frac{GM(h)}{\az h^2}\right] \frac{M(h)}{2\pi h^2},  \eeqno{dynda}
\beq \SD\^{0,sph}= -\frac{1}{\tpg}\int_0^\infty\div\{\n\left[ \frac{GM(z+h)}{\az(z+h)^2} \right]\vgN\}~dz,             \eeqno{dyndasph}
from which
\beq \SD\^{0,sph}=2\int_0^\infty\n\left[ \frac{GM(z+h)}{\az(z+h)^2} \right]\r_s(z+h)~dz+\frac{1}{\tpg}\int_0^\infty\gN\frac{d\n}{dz}dz            \eeqno{dyndasphaa}
($\vgN$  points to the origin along the $z$ axis, and $\gN=|\vgN|$),
and after some manipulations (change of variables from $z$ to $Y=\gN/\az$, and integration by parts, with $\bar z=z+h$)
\beq \Sdz=\SD\^{0,disc}+\SD\^{0,sph}=
2\int_h^\infty\n\left[ \frac{GM(\bar z)}{\az\bar z^2} \right]\r_s(\bar z)~d\bar z+\SM\int_{0}^{\gN^+(0)/\az}\n(Y)dY. \eeqno{masmer}
However, in the last integral, $Y=\gN(0,z)/\az=GM(\bar z)/\az\bar z^2$.
So, changing variables from $Y$ to $\bar z$,
\beq dY=d\bar z\left[\frac{-2GM(\bar z)}{\az\bar z^3}+\frac{\fpg\r_s(\bar z)}{\az}\right]. \eeqno{kalao}
Putting all the above together we get
\beq  \Sdz=\frac{1}{\pi}\int_h^\infty \n\left[\frac{GM(\bar z)}{\az\bar z^2}\right]\frac{M(\bar z)}{\bar z^3}d\bar z.  \eeqno{dbadba}
\par
A pure Kuzmin disc is gotten when $M(\bar z\ge h)=M_k$ is constant. Then, change variables in the integral to $Y=GM_K/\az\bar z^2$ to obtain
\beq  \Sdz=\SM\int_0^{\Sbz/\SM} \n(Y)dY=\SM\SS(\Sbz/\SM).  \eeqno{dbazada}
This is the universal expression predicted by AQUAL/QUMOND; here we see that it is predicted for Kuzmin discs in the general class of MG theories.
\par
The radial MOND acceleration in the plane of the disc is
\beq g_r(r)=g(\bar R)\frac{r}{\bar R}=\n\left[\frac{GM(\bar R)}{\az\bar R^2}\right]\frac{GM(\bar R)r}{\bar R^3},  \eeqno{grara}
from which we get the acceleration integral
\beq  \G =\int_0^\infty \frac{g(\bar R)}{\bar R}dr=\int_h^\infty \frac{g(\bar R)}{(\bar R^2-h^2)^{1/2}}d\bar R.  \eeqno{jijava}
\par
For completeness, we can also calculate the value of $\G$ for these models, with the MI expression, whereby the accelerations are determined from rotational speeds of exactly circular trajectories. In this case, the MOND radial acceleration is related to the Newtonian radial acceleration by
$g_r\^{MI}=\n(g_r\^N/\az)g_r\^N$ instead of $g_r\^{MG}=\n(g\^N/\az)g_r\^N$ (the difference being that in the former the $r$ component appears in the argument of $\n$ instead of the full Newtonian acceleration in the latter).
We are now equipped to examine the correlations we are after.
\subsubsection{The $\Sdz-\G$ correlation  \label{kuzrel}}
Using expression (\ref{dbadba}) for $\Sdz$ and expression (\ref{jijava}) for $\G$, we get
\beq \k=\frac{\tpg\Sdz}{\G}=2\int_h^\infty \frac{g(\bar R)~d\bar R}{\bar R}\left[\int_h^\infty \frac{g(\bar R)~d\bar R}{(\bar R^2-h^2)^{1/2}}\right]^{-1}.   \eeqno{mimili}
The two integrals differ only by the replacement of one power of $\bar R$ in one, by $(\bar R^2-h^2)^{1/2}$ in the other.
For $h=0$, for which the model is spherical, we get $\k=2$ as expected. For $h>0$, the second integral is larger than the first and $k<2$.
\par
For a Kuzmin disc, for which $M(\bar R>h)=M\_K$ constant, we have
\beq \k=2\int_h^\infty \n\left[\frac{GM\_K}{\az\bar R^2}\right]\frac{d\bar R}{\bar R^3}\left\{\int_h^\infty \n\left[\frac{GM\_K}{\az\bar R^2}\right]\frac{d\bar R}{\bar R^2(\bar R^2-h^2)^{1/2}}\right\}^{-1}.   \eeqno{mimika}
In the Newtonian limit, $\n\equiv 1$, one gets $\k=1$, as expected from Toomre's result. However, in MOND, the dynamical mass acquires a phantom bulge, so we expect $\k$ to be between 1 and 2. Indeed, the $\n$ factor in the integrands increases with increasing values of $\bar R$. This causes the value of $\k$ to increase over its value for $\n=1$. For example, in the DML, where we can put $\n(Y)=Y^{-1/2}$, we get from expression (\ref{mimika}) $\k=4/\pi$.
\subsubsection{The $\Sdz-\Sbz$ correlation}
We have several expressions for the $\Sdz$ and the corresponding ones for $\Sbz$, gotten from them by putting $\n(Y)\equiv 1$. Such pairs can be used to asses the level of the CSDC in our class of theories. All these theories (including AQUAL and QUMOND) give the same results for the Kuzmin-like, disc-plus-bulge mass models, when expressed in terms of the one-dimensional interpolating function, $\n(Y)$.
\par
Take, for example, the pair $\Sbz$ from Eq. (\ref{bzaza}) (second equality) and $\Sdz$ from Eq. (\ref{dbadba}). As expected, there is no universal
functional relation between the two, as there is for AQUAL and QUMOND for pure discs. We already saw that for a spherical mass the correlation does depend on the mass distribution.
\par
As regards the asymptotes, the Newtonian limit takes the form that it always does, which
$\Spz=\Sdz-\Sbz\rar \a\SM$, with $\a$ depending only on the exact form of $\n(Y)$, as given in Eq. (\ref{matesa}).
In the DML, we get the value of $\eta\equiv\Sdz/(\SM\Sbz)^{1/2}$ from Eqs. (\ref{bzaza}) and (\ref{dbadba}),
\beq  \eta=\sqrt{2}\frac{\int_h^\infty \bar z^{-2}M^{1/2}(\bar z)~d\bar z}{\left[\int_h^\infty \bar z^{-3}M(\bar z)~d\bar z\right]^{1/2}}.   \eeqno{magasha}
As predicted by the basic tenets, we see that $\eta$ does not depend on the normalization of $M$, nor on the characteristic scale lengths of the mass distribution, as long as we scale them and $h$ by the same factor.
For $M(R)$ constant beyond $R=h$, we have the Kuzmin pure disc, and expression (\ref{magasha}) gives $\eta=2$. For $h=0$ we get the result for the spherical case, Eq. (\ref{liouter}). For AQUAL/QUMOND this is the value that applies to all pure discs; here we derived it fora Kuzmin disc in the general class of theories.

\subsection{The CSDC for pure discs  \label{puredisc}}
We saw that AQUAL and QUMOND predict a universal -- i.e., disc independent -- $\Sbz-\Sdz$ relation for pure discs, the relation hinging only on the $\n$ interpolating function of these theories. We also saw that the more general class of MG theories predict the same relation, but only for Kuzmin discs, in which the {\it one-dimensional} interpolating function of the specific theory appears in the same way.
A Kuzmin disc is a mass model whose gravitational field is describable, in some sense, in terms of a one-dimensional mass model, so perhaps this is not so surprising.
But, in the general class of theories, it cannot be expected that a similar strict algebraic relation holds for general pure discs and, in particular, that only the one-dimensional interpolating function will define a CSDC for such discs. What can we, nevertheless, say about such a CSDC?
\par
One general result that applies to all the theories in the class discussed in Sec. \ref{genmg}
is that there appears a phantom disc, whose local surface density, $\S\_P^{disc}$ depends only on the local baryonic disc surface density, $\S\_B^{disc}$, according to
\beq \S\_P^{disc}=\left[\n\left(\frac{\S\_B^{disc}}{\SM}\right)-1\right]\S\_B^{disc}, \eeqno{numara}
where $\n(Y)$ is the one-dimensional interpolating function of the theory, which is derived from Eqs. (\ref{eqfarad}) (\ref{eqpsarad}). This results from the fact that the jump conditions of the MOND and the Newtonian fields across a thin disc are related by the one-dimensional behavior of the theory. Relation (\ref{numara}) holds, in particular, for the relation between the central surface densities $\S\_P^{0,disc}$ and $\S\_B^{disc}$.\footnote{Such a phantom disc also occurs when there is a bulge present.}$^,$\footnote{Note that this is a secondary prediction of MOND that holds in this class of MG theories, but it does not follow from the basic tenets. In Ref. \cite{milgrom22a}, I showed that MI theories, for example, can make rather different predictions here. In particular, the notion of phantom disc is not defined in such theories.}
\par
There is no baryonic mass outside the disc, so there we have the contribution to the dynamical, central surface density from a phantom bulge
\beq \Sdz(out)=\frac{1}{\tpg}\int_{0^+}^{\infty} \Delta\f~d\z,\eeqno{xi}
where $\f$ is the MOND potential gotten by solving Eqs. (\ref{eqfa}), and the integration is along the $z$ axis from $\z=0^+$ just outside the disc. ({\it Here and hereafter, I use $\z$ for the $z$ coordinate, to avoid confusion with the above use of $z$ as a scalar variable in the TRIMOND Lagrangian.})
\par
The formulation of the general Lagrangian is not specific enough for me to derive further general results, relevant to the CSDC; so I will confine myself to the subclass of TRIMOND theories, where the MOND potential enters the gravitational Lagrangian in a specific manner, as explained in Sec. \ref{trimond}.
\subsubsection{The pure-disc CSDC in TRIMOND}
In TRIMOND theories, $\f$ in Eq. (\ref{xi}) is to be determined from  the TRIMOND field equations (\ref{vvvf}) and (\ref{mail}).
Given that the Newtonian field $\psi$ satisfies $\Delta\psi=0$ outside the disc, we can get from Eq. (\ref{vvvf}),
\beq   \Delta\vrf=-(\grad\psi\cdot\grad\F_z+\grad\vrf\cdot\grad\F_y)/\F_y.   \eeqno{mm}
Substituting in Eq. (\ref{mail}) we get an expression for $\Delta\f$
\beq  \Delta\f=\grad\psi\cdot\grad\F_x+\grad\vrf\cdot\grad\F_z-\frac{\F_z}{\F_y}(\grad\psi\cdot\grad\F_z+\grad\vrf\cdot\grad\F_y).   \eeqno{mm1}
In the integrand in Eq. (\ref{xi}) $\Delta\f$ is needed only on the $\z$ axis, where, from the symmetry, all the gradients are along the axis, so we can write
\beq \Sdz(out)=\frac{1}{\tpg}\int_{0^+}^{\infty}d\z~ \left(\frac{d\F_x}{d\z}-\frac{\F_z}{\F_y}\frac{d\F_z}{d\z}\right)g\_N
+\left(\frac{d\F_z}{d\z}-\frac{\F_z}{\F_y}\frac{d\F_y}{d\z}\right)g\_{\vrf}.\eeqno{mm2}
This expression is useful because the integrand now depends only on the accelerations on the $\z$ axis, since all the second derivatives have disappeared. On the $\z$ axis, $\F_x,~\F_y,~\F_z$ are also functions of only $\gN$ and $g\_{\vrf}$.
\par
Near the disc and far from it -- more specifically, at $\z$ values small and large compared with the central, characteristic size of the disc -- the gravitational field becomes one dimensional, so the algebraic relations (\ref{mumuiop}) and (\ref{mumulig}) hold. Had this been the case all along the $\z$ axis, we would have gotten the same relation (involving the one-dimensional interpolating function of the theory) between $\Sdz$ and $\Sbz$ as we have for Kuzmin discs, where, indeed, the field outside the disc is one dimensional. Thus, the deviation from such a relation, and the dependence on the specific disc model, result from the contribution of the intermediate region on the $\z$ axis.
\par
My aim here is, indeed, to demonstrate in more detail why, in general, the relation between $\Sdz$ and $\Sbz$ for pure discs is not universal, as in the special cases of AQUAL and QUMOND. For this, it is enough to further confine ourselves to TRIMOND theories in which $\F_z=\eps$ is a constant (AQUAL and QUMOND are such cases). Equation (\ref{mm2}) then reduces to
\beq \Sdz(out)=\frac{1}{\tpg}\int_{0^+}^{\infty} \left (\frac{d\F_x}{d\z}\gN  -\frac{\eps}{\F_y}\frac{d\F_y}{d\z}g\_{\vrf}\right)~d\z. \eeqno{vii}
The field configuration near the disc (e.g., at $0^+$) is one dimensional and the three accelerations are related there
 by the one-dimensional relations (\ref{mumuiop}) and (\ref{mumulig}) reduced to our special case,
\beq  \F_y^+ g\_\vrf^++\eps g\_N^+=0,   \eeqno{teriop}
\beq  g^+=\F_x^+ g\_N^++\eps g\_\vrf^+.    \eeqno{terlig}
\par
Since the only quantities that enter the integrand in Eq. (\ref{vii}) are the acceleration values on the $\z$ axis,
{\it for a given disc}, we can think of one quantity in the integrand as a function of any other. For example, along the $\z$ axis, $\F_x(\z)$ can be treated as a function $\n\_1(Y)$, where $Y\equiv g\_N$ (in this section, again, we put $\az=1$ for convenience). Strictly speaking this is only so if $Y(z)$ is monotonic from its value $\gN^+$ to $0$ at infinity, because then $Y(\z)$ can be inverted and $\z(Y)$ gives $\F[\z(Y)]$. There are discs for which this is not so; e.g., for a disc with a hole at the center, $\gN$ vanishes at the origin and at infinity. In this case we need to make the treatment piecewise. But here, for the demonstration, we deal with discs for which inversion is possible.
\par
$\n\_1[Y(\z)]$ is not universal for the given theory, like the one-dimensional interpolating function, depending not only on the disc model, but also on the value of $\Sbz$ relative to $\SM$; it is defined only on the $\z$ axis, and only for argument values between $0$ and $\gN^+$.
\par
Write, then, the integral of the first term in Eq. (\ref{vii})
\beq \int_{0^+}^{\infty} \left [\frac{d}{d\z}[Y\n\_1(Y)]-\n\_1(Y)\frac{dY}{d\z}\right]d\z= -[Y\n\_1(Y)]\^{0^+}+\int\^{Y(0^+)}\_{Y(\infty)} \n\_1(Y)dY=-g\_N\^+\n\_1(g\_N\^+)+\int_0\^{g\_N\^+}\n\_1(Y)~dY.   \eeqno{xx}
\par
In a similar vein, on the $\z$ axis, we can think of $\F_y$ as a function of $g\_{\vrf}(\z)$: $F_y(\z)=\m[g\_\vrf(\z)]$.
Then, if we denote
\beq Y=-\eps^{-1} X\m(X),   \eeqno{x}
 and invert this relation, then $\n\_2(Y)$
is defined such that
\beq X=\eps^{-1}Y\n\_2(Y).   \eeqno{xv}
Take the $X$ derivative of Eq. (\ref{x}) to get
\beq  -\frac{\eps}{\m}\frac{dY}{dX}-1=\frac{X\m'(X)}{\m}.  \eeqno{xii}
But, from Eqs. (\ref{x}) and (\ref{xv})
 we have
\beq  1/\m=-\eps^{-2}\n\_2,   \eeqno{xara}
 and the second integrand in Eq. (\ref{vii}) is $(-\eps X/\m)\m'(dX/d\z)$.
Then, the second integral can be written as
\beq -\eps X(0^+)+\int_0^{g\_N^+}\n\_2(Y)~dY.  \eeqno{xiii}
where we used the fact that at $0^+$ the one-dimensional relations hold; so, for example, $Y(0^+)=-\eps^{-1}X\m(X)(0^+)=g\_N^+$ [from Eq. (\ref{teriop})].
All in all, we get
\beq \Sdz(out)=\frac{1}{\tpg}\{-\F_x^+g\_N^+-\eps X^+   +\int_0^{g\_N^+}[\n\_1(Y)+\n\_2(Y)]dY\}.\eeqno{xiv}
From Eq. (\ref{terlig}) the first  two terms give $-(\tpg)^{-1}g^+=-\Sdz(disc)$, which cancels with the contribution of the disc to $\Sdz$, to give (restoring $\az$)
\beq \Sdz=\SM\int_0^{\Sbz/\SM}\n^*(Y)~dY\equiv \SM\SS^*(\Sbz/\SM), \eeqno{xvii}
where $\n^*(Y)=\n\_1(Y)+\n\_2(Y)$. This is similar in form to the universal AQUAL/QUMOND relation (\ref{vicha}), but here $\SS^*$ is not universal. In fact, being too dependent on the model, $\n^*(Y)$ is not very useful for calculations, as it is not known beforehand for a given theory and model. I use this concept here only to better bring out the differences from AQUAL and QUMOND.
\par
Indeed, in the above derivation, we encounter, in a sense, a combination of AQUAL and QUMOND. Remember that QUMOND is a special case with $\eps=0$, so $\n_2=0$, and $\n^*=\n_1$ is then the interpolating function of QUMOND. For AQUAL, $\F_x\equiv 0$; so, $\n_1=0$, and $\n^*=\n_2$ is the AQUAL, $\n$-type interpolating function.
\par
Also, $\n^*$ here has the same asymptotic limits as $\n$.
For $Y\gg 1$, the statement around Eq. (\ref{newman}), on the Newtonian limit of TRIMOND, tells us that
in this limit $\n\_1=\F_x\rar\b$, and from Eq. (\ref{xara}) and the definition of $\m=\F_y$, $\n\_2\rar -\eps^2/\o$. So, by Eq. (\ref{newman}), $\n\_1+\n\_2\rar 1$.
\subsubsection{The deep-MOND limit of TRIMOND theories with $\F_z$ constant  \label{fzcon}}
I now describe in some detail the form that the DML equations take for this subclass, also as an opportunity to see how the DML in TRIMOND theories can differ from that in AQUAL/QUMOND, the Lagrangian of which involves a function of a single variable. The DML in TRIMOND, whose Lagrangian involves generally a function of three variables was discussed in Ref. \cite{milgrom23}.
\par
The Lagrangian of the subclass I consider here involves a function of two variables, and has the form
\beq  \F(x,y,z)=\eps z+\bar\F(x,y).  \eeqno{mazepo}
From the requirement of scale invariance in the DML, expressed as Eq. (\ref{siff}), the DML of $\F$ has to be such that
\beq \eps z+\bar\F(x,y)=\l^3[\eps \l^{\a-3}z+\bar\F(\l^{-4}x,\l^{\a-2}y)],    \eeqno{mapo}
for some $\a$ (which is the scaling dimension of $\vrf$) and for any $\l>0$. This means that we have to take $\a=0$, and the DML of $\bar\F$ has to satisfy
\beq \bar \F(x,y)=\l^3\bar\F(\l^{-4}x,\l^{-2}y).   \eeqno{mishte}
This implies that $\bar\F$ is of the form (taking $\l=x^{1/4}$)
\beq \bar\F(x,y)=x^{3/4}F(y/x^{1/2}).  \eeqno{nutov}
Any choice of $F$ would give a scale invariant DML, pending some consistency requirement, such as existence and uniqueness of solutions. In contradistinction from AQUAL and QUMOND, where the DML of the theory is completely fixed by the requirement of scale invariance, here one still needs to specify a function of one variable to specify the theory's DML (for general TRIMOND theories, one is left with the choice of a function of two variables).
From Eq. (\ref{nutov}) follows that in the DML,
\beq  \F_x=\frac{3}{4}x^{-1/4}F-\oot yx^{-3/4}F',~~~~~~\F_y=x^{1/4}F', ~~~~~~~~\F_z=\eps.   \eeqno{allder}
The standard choice of the normalization of $\az$ is such that asymptotically far from a mass, we have $g=\sqrt{\az\gN}$. This normalization still has to be imposed, as follows:
In the spherical case, which is relevant to this, we use the algebraic relation (\ref{mumuiop}), which reads here
\beq  F'(q)q^{1/2}+\eps=0,~~~~~q\equiv yx^{-1/2}.    \eeqno{nukvar}
This implies that in a one-dimensional, DML system, $y=q_0x^{1/2}$, where $q_0$ is the solution of Eq.(\ref{nukvar}).
Thus, for the theory to be consistent, $F'(q)$ has to be such that this equation has a solution, and that the solution is unique.
\par
The second one-dimensional equation (\ref{mumulig}) then tells us that
\beq g=\frac{3}{4}[F(q_0)-2q_0F'(q_0)]\gN^{1/2}.   \eeqno{gamarnu}
We want to choose the still-free additive constant to $F$, so that in this case $g=\gN^{1/2}$ ($\az=1$).
\par
In summary, in constructing such a theory, we choose $F$ up to an additive constant, and a value of $\eps$. Then, Eq. (\ref{nukvar}) determines $q_0$, and Eq. (\ref{gamarnu}) determines the additive constant.
\par
While, from their definition, $\n_1(Y)$, and $\n_2(Y)$, and hence $\n^*(Y)$, scale as $Y^{-1/2}$ for $Y\rar 0$, the proportionality constant depends on the disc model. While in AQUAL/QUMOND, the canonical normalization of $\az$ determines the DML normalization $\n(Y)=Y^{-1/2}$, here it does not determine that of the DML of $\n^*$, and hence of the value of $\eta$.

\section{Summary \label{summary}}
The CSDC is a correlation between the dynamical central surface density, $\Sdz$, of an axisymmetric and plane-symmetric, galactic system, and the baryonic central surface density, $\Sbz$.
Various versions of the CSDC emerge from various MOND starting points.
\par
I have emphasized, in the first place, that the notions of dynamical density and dynamical surface density are defined only in modified-gravity theories. In these, the Newtonian potential is modified into a MOND potential, from which one can define a dynamical density: the density that  would source the MOND potential in a Poisson equation.
In modified-inertia versions of MOND there is no gravitational potential defined, and so no dynamical density is defined. However, since in actual tests one uses proxies for the dynamical surface density, and these may be well defined in MI theories, the CSDC may emerge in the MI context in some circumstances.
\par
I showed that from MOND's defining tenets alone follows a CSDC in a basic form: The underlying tenets pertain to the behavior of the theory at high and low accelerations, but do not specify the interpolation scheme between these behaviors, and how it takes place in different phenomena.\footnote{The only assumption about the interpolation is that it occurs within a range of order $\az$ around $\az$; so that the theory does not involve dimensionless constants $\gg 1$ or $\ll 1$.} So, they can predict only the high- and low-surface-density ends of the CSDC.
\par
At the high end, $\Sbz\gg\SM$, where $\SM$ is the MOND critical surface density, it is predicted that the phantom surface density
$\Spz\equiv\Sdz-\Sbz\rar \c\SM$, with $\c=O(1)$ that depends on the theory, but not on the system, because in this limit the contribution to $\Spz$ comes from far outside the system, where it can be thought of as a point mass, whose structure is immaterial here.
\par
At the other end, $\Sbz\ll\SM$, it is predicted that $\Sdz=\eta(\SM\Sbz)^{1/2}$, with $\eta= O(1)$ that does not depend on the mass or overall size of the system, but that can depend on dimensionless attributes of the system, such as whether we have a sphere, a pure disc, or a combination, or generally on ratios of scale-length attributes of the system.
\par
{\it This makes a rough CSDC, as described above, a primary prediction of MOND that is shared by all MOND theories that embody the basic axioms.}
\par
Specific MOND theories can predict more constrained forms of a CSDC, especially if we also consider specific system geometries.
For example, the AQUAL and QUMOND theories predict for pure discs a universal (disc-independent), tight algebraic relation $\Sdz(\Sbz)$.
\par
As other examples, I considered the detailed predictions of a large class of modified-gravity theories that include AQUAL, QUMOND, and TRIMOND as special cases. They all coincide in their predictions for one-dimensional systems, such as spherical systems or near-disc dynamics. Such predictions then hinge only on the so-called one-dimensional interpolating function -- a function of the Newtonian acceleration of the system.
\par
Importantly, though, and in contradistinction from the special cases of AQUAL and QUMOND, these theories do not predict a universal relation for pure discs, with the exception of the Kuzmin disc for which they all make the same prediction of a tight algebraic CSDC.
In particular, the value of $\eta$ that describes the low-$\Sbz$ limit can depend somewhat on the exact disc model, and on the exact theory.
Such details of the CSDC are thus secondary MOND predictions, as they do not follow from the basic tenets alone.

\appendix
\section{Relation between $\Sdz$ and the acceleration integral $\G$ for some mass models  \label{appendixa}}
Here I derive the relation between central surface density and the acceleration integral for some mass models.
To avoid confusion, note that because this relation is linear, and is derived on the basis of Newtonian dynamics, these models model not the baryonic mass distribution of the galaxy, but its MOND, dynamical mass distribution (baryonic plus phantom). Hence these surface densities are designated with a subscript D.
For example, a pure baryonic disc, has both a disc and a spheroidal (``bulge'') component of dynamical mass; so the Miyamoto-Nagai models below
represent this compound distribution.
\subsection{Spheroids  \label{appendix1}}
Consider a thin spherical shell of radius $r_s$ and surface density $\S_s$, so its mass is $M_s=4\pi r_s\^2\S_s$.
The (Newtonian) acceleration vanishes inside the shell, and outside it is $g(r)=GM_s/r\^2$.
Its central surface density is $2\S_s$, and the acceleration integral is
\beq \G\equiv\int_0^\infty g~d\ln{r}=\frac{GM_s}{2r_s\^2}=\tpg\S_s,  \eeqno{a1}
so it has $\k=2$, as expected.
\par
Contracting or expanding the shell by a factor $q$ along the $z$ axis, we get a homoeoid thin shell, which still produces zero acceleration inside it.
The acceleration along the equator, a distance $r>r_s$ from the center is
$g(r)=\S_sf(q,r_s,r)$, but on dimensional grounds it has to be of the form
\beq   g(r)=\fpg \S_s \hat f(q,r/r_s).  \eeqno{a2}
The acceleration integral is thus
\beq \G=\fpg\S_s \int_{r_s}^{\infty}\hat f(q,r/r_s)d\ln{r}=\fpg\S_s \int_{1}^{\infty}\hat f(q,x)d\ln{x}\equiv \fpg\S_s/\k(q).  \eeqno{a3}
We see then that for this shell, $\k$ defined in Eq. (\ref{iva}) is $\k(q)$, which depends only on $q$.
\par
Because of the additivity of the central surface density and of $\G$, the same relation holds for any spheroid of $r$-independent oblateness factor $q$.
\par
Clearly, $\k(q)$ is a decreasing function of $q$: the smaller $q$ is the nearer the shell elements are to a point on the external equator, and, in addition, the larger the radial component is. So, $\G$ increases with decreasing $q$, but $\Sdz$ is independent of $q$, so $\k$ decreases.
\par
To get an explicit expression for $k(q)$, start from Eq. (2-91) in Ref. \cite{bt87}  (somewhat paraphrased), which gives the acceleration in the midplane for an axisymmetric spheroid, of density distribution (in cylindrical coordinates $r,z$
\beq \r(r,z)=\hat\r(y),~~~~~~y\equiv \left(r^2+\frac{z^2}{q^2}\right)^{1/2}, \eeqno{a4}
as
\beq g(r)=\frac{\fpg q}{r}\int_0^r\frac{\hat\r(y)y^2dy}{\sqrt{r^2-e^2y^2}},~~~~~e=\sqrt{1-q^2}.   \eeqno{a5}
The acceleration integral is then
\beq \G=\fpg q\int_0^\infty \frac{dr}{r^2}\int_0^r\frac{\hat\r(y)y^2dy}{\sqrt{r^2-e^2y^2}},   \eeqno{a6}
which after some rearrangement and changing the order of the integration gives
\beq    \G=\fpg q\int_0^1\frac{x~dx}{\sqrt{1-e^2x^2}}     \int_0^\infty \hat\r(u)du=\frac{\fpg q}{1+q}\int_0^\infty \hat\r(u)du. \eeqno{a7}
The central surface density is
\beq \Sdz=2\int_0^\infty \r(0,z)dz=2q\int_0^\infty \hat\r(u)du;  \eeqno{a8}
Putting the two together we get from the definition of $\k$, Eq. (\ref{iva}),
\beq  \k(q)=1+q.   \eeqno{a9}
\subsection{Miyamoto-Nagai models   \label{appendix2}}
For Miyamoto-Nagai models the potential field is
\beq  \f(r,z)=-\frac{MG}{\{r^2+[a+(z^2+b^2)^{1/2}]^2\}^{1/2}}.   \eeqno{marqui}
So, in the symmetry plane $g_r(r,0)=MGr/[r^2+(a+b)^2)]^{3/2}$ yielding
\beq \G= \frac{MG}{(a+b)^2}   .   \eeqno{lalalai}
The source density along the $z$ axis is (e.g., Ref. \cite{bt87})
\beq  \r(0,z)= \left(\frac{Mb^2}{4\pi}\right)\frac{a+3(z^2+b^2)^{1/2}}{[a+(z^2+b^2)^{1/2}]^3(z^2+b^2)^{3/2}},      \eeqno{matatah}
and the expression for $\Sdz$ can be brought to the form
\beq  \Sdz=  \left(\frac{M}{2\pi b^2}\right) \int_0^1\frac{x^3(3+\l x)dx}{(1+\l x)^3\sqrt{1-x^2}},    \eeqno{minaga}
from which
\beq \k=(1+\l)^2\int_0^1\frac{x^3(3+\l x)dx}{(1+\l x)^3\sqrt{1-x^2}},  \eeqno{nanmip}
where $\l=a/b$. As $\l$ goes from $0$ (Plummer sphere) to $\infty$ (Kuzmin disc), $\k$ goes, monotonically, from 2 to 1.

%\clearpage
\end{document}